\documentstyle[pre,aps,preprint,eqsecnum,epsfig]{revtex} 

\tightenlines

\newcommand{\app}{\rightarrow}
\newcommand{\bigi}{I}

\newcommand{\df}{\Delta F}
\newcommand{\dfnp}{{\cal F}}

\newcommand{\ham}{{\cal H}}

\newcommand{\ihat}{\hat{I}}
\newcommand{\idag}{I^\dagger}
\newcommand{\iddag}{I^\ddag}
\newcommand{\kap}{\hat{\kappa}}
\newcommand{\levy}{L\'{e}vy}

\newcommand{\muz}{\hat{\mu}}
\newcommand{\muq}{\tilde{\mu}}
\newcommand{\phiq}{\varphi^{(q)}}
\newcommand{\rmpq}{R^{(q)}}

\newcommand{\sigz}{\hat{\sigma}}
\newcommand{\sigq}{\tilde{\sigma}}
\newcommand{\totd}{{\mathrm{d}}}
\newcommand{\xbf}{\textbf{x}}

\newcommand{\xavg}{\langle X \rangle}
\newcommand{\xhat}{\hat{x}}
\newcommand{\xnl}{\xavg^{g}}

\begin{document}
\title{Systematic Finite-Sampling Inaccuracy in Free Energy Differences and Other Nonlinear Quantities}
\author{Daniel M. Zuckerman$^{1,2}$ and Thomas B. Woolf$^{3,4}$}
\address{
$^1$Center for Computational Biology \& Bioinformatics, \\ 
University of Pittsburgh, 200 Lothrop Street, Pittsburgh, PA 15261; \\
$^2$Department of Environmental \& Occupational Health, \\
Graduate School of Public Health, University of Pittsburgh; \\
$^3$Department of Physiology and  $^4$Department of Biophysics,\\
Johns Hopkins University School of Medicine, Baltimore, MD 21205\\
\texttt{dzuckerman@ceoh.pitt.edu, woolf@groucho.med.jhmi.edu}
}

\date{to appear in the Journal of Statistical Physics}
\maketitle

\begin{abstract}
Systematic inaccuracy is inherent in any computational estimate of a non-linear average, such as the free energy difference $\df$ between two states or systems, because of the availability of only a finite number of data values, $N$.
In previous work, we outlined the fundamental statistical description of this ``finite-sampling error.''
We now give a more complete presentation of 
(i) rigorous \emph{general} bounds on the free energy and other nonlinear averages, which generalize Jensen's inequality;
\hbox{(ii) asymptotic} $N \app \infty$ expansions of the average behavior of the finite-sampling error in $\df$ estimates;
(iii) illustrative examples of large-$N$ behavior, both in free-energy and other calculations; and
(iv) the universal, large-$N$ relation between the average finite-sampling error and the fluctuation in the error.
An explicit role is played by {\levy } and Gaussian limiting distributions.
\end{abstract}


\newpage
\section{Introduction}

Because of the substantial recent interest in free energy difference $\df$ calculations 
(e.g., \cite{Kollman-1994,Jarzynski-1997a,Jarzynski-1997b,Bruce-1997,Kofke-1999,Kofke-2001a,Kofke-2001b,deKoning-1999,Hummer-2001b,Jarzynski-2001b,Schulten-2001,Bustamante-2002,Zuckerman-2002a,Zuckerman-2002b}), this report discusses the unavoidable error that arises from use of a finite amount of computer time.
There is a tremendous range of applications for computational $\df$ estimates in physical, chemical, and biological systems.
Examples include computations relating crystalline lattices \cite{Bruce-1997,deKoning-1999}, the behavior of magnetic models \cite{deKoning-1999,Reinhardt-1992}, and biomolecular binding events --- of ligands to both DNA and proteins (e.g., \cite{McCammon-1984,Beveridge-1989,McCammon-1991,Kollman-1993,Kollman-1994}).
%
Computations of $\df$, moreover, are formally equivalent to calculating the temperature dependence $F(T)$ \cite{deKoning-1999}.
Most recently, it has been pointed out that $\df$ calculations are required to convert \emph{experimental} data from nonequilibrium single-molecule pulling measurements to free energy vs.\ extension profiles \cite{Hummer-2001b,Bustamante-2002}; see also \cite{Schulten-2001}.
From a methodological standpoint, free energy computation protocols have been the subject of long and sustained interest \cite{Widom-1963,Valleau-1972,Bennett-1976,McCammon-1984,Jorgensen-1985,Cross-1986,Reinhardt-1990,Wood-1991,Hermans-1991,vanGunsteren-1991,Mezei-1993,Brunger-1993,Hummer-1996,Schon-1996,Brooks-1996,Tidor-1997,Kofke-1999,Reinhardt-2000,Roux-2000,Hummer-2001,Kofke-2001a,Kofke-2001b,vanGunsteren-2001,Hermans-2002}

Free energy computations have long been recognized to suffer from a number of sources of error: (i) inaccuracy of the model (Hamiltonian) itself, (ii) incomplete conformational sampling, and (iii) finite sample size.
In biomolecular systems, the issue of model accuracy (i) is indeed important, as typical all-atom force-fields are generally assumed to be capable of no more than 1 - 2 $k_B T$ accuracy in estimates of free energies --- e.g., of ligand binding.
That is, even with perfect sampling, computational estimates typically will not match experimental values.
Second, like every simulation technique, free energy calculations are subject to errors based on (ii) incomplete conformational sampling.
``\emph{Incomplete}'' sampling here refers to the lack of access to important parts of phase or conformational space: that is, the distribution of samples of size $N$ generated by the simulation will \emph{not} match the true (representative) distribution of size-$N$ samples which would be drawn at random from a very long, perfectly-sampled simulation.  
Incomplete conformational sampling introduces bias into even simple computations attempting to estimate linear quantities, such as the mean of some coordinate or function.
Conformational sampling errors in free energy calculations have long been recognized as ``hysteresis'' or ``Hamiltonian-lag,''
and a number of workers have made important contributions toward understanding and overcoming these errors --- e.g., \cite{Wood-1991,Hermans-1991,Brunger-1993}.

We focus here on the third type of error (iii) that due \emph{solely} to the necessarily finite samples collected in a simulation.
Such finite-sampling bias occurs in every non-linear calculation, as detailed below, and should be clearly distinguished from the independent error due to (ii) inadequate conformational sampling.
Specifically, finite-sampling error occurs even when conformational sampling is perfect --- i.e., when representative samples of finite size are generated by the simulation.
Finite-sampling errors in computational estimates of $\df$ were first recognized by Wood and coworkers \cite{Wood-1991b} and later discussed by others \cite{Jarzynski-1997b,Kofke-2001a,Kofke-2001b,Zuckerman-2002a,Zuckerman-2002b};
see also \cite{Stone-1982,Ferrenberg-1991,Landau-Binder}.
The in-depth work of Lu and Kofke presents an entropy-based description of finite-sampling errors, which emphasizes the critical asymmetry between generalized ``insertion'' and ``deletion'' calculations \cite{Kofke-2001a,Kofke-2001b}. 

Figure \ref{fig:examples} illustrates the phenomenon of finite-sampling error for a mathematical model and for a biological system \cite{Nanda-FA}, emphasizing the universality of finite-sampling errors.
Because these inaccuracies can be many times $k_B T$ (see Fig.\ \ref{fig:examples} and Ref.\ \cite{Zuckerman-2002a}) --- especially in the important context of biomolecular calculations where large system sizes limit the quantity of data available for analysis --- there is a strong motivation to understand and overcome these errors.
Ferrenberg, Landau and Binder showed that finite-sampling errors accompanying susceptibility computations can be understood on the basis of elementary statistical principles \cite{Ferrenberg-1991}; 
however, the errors in \emph{non-linear averages} like $\df$ apparently had remained without an explicit theoretical basis until recently \cite{Kofke-2001a,Kofke-2001b,Zuckerman-2002b}.
In a recent monograph, in fact, Landau and Binder note that finite-sampling errors are ``generally given inadequate attention'' \cite{Landau-Binder}.

This report both provides fuller details of the theory outlined in \cite{Zuckerman-2002b}, and also presents new results.
Our report includes 
(i) a detailed proof that the expected value of a finite-data $\df$ estimate ($\df_n$) bounds the true free energy --- \emph{independent of the distribution of underlying work values}; 
(ii) full derivations of the asymptotic expressions for $\df_n$ as $n\app\infty$ for arbitrary --- including long-tailed --- distributions of the work ($W$) data used to estimate $\df$;
(iii) analogous derivations for the root-mean-square and related ``geometric'' non-linear averages;
(iv) derivation and numerical demonstration of the universal asymptotic relation between $\df_n$ and its fluctuation.
As in our brief report \cite{Zuckerman-2002b}, the present discussion makes use of mathematical results regarding the convergence --- to ``stable'' limiting distributions \cite{Feller-1971,Zolotarev-1986,Uchaikin-Zolotarev}, also known as \levy\ processes (e.g., \cite{Shlesinger-1995}) --- of the distributions of sums of variables.

In outline, the paper now proceeds to Sec.\ \ref{sec:formal} where formal groundwork for the discussion is laid.
Section \ref{sec:bound} rigorously proves the true free energy $\df$ is bounded by $\df_n$, the expected value of a free-energy estimate based on a finite quantity of data; 
analogous bounds apply for \emph{arbitrary} non-linear averages.
Derivations of the asymptotic series for $\df_n$ are given in sections \ref{sec:dfn-finite} and \ref{sec:dfn-diverge},
while Section \ref{sec:fluct} derives the universal relation between $\df_n$ and its fluctuation.
We conclude with a summary and discussion of the results in Section \ref{sec:conclude}.

\section{Free-Energy Estimates from Finite Samples}
\label{sec:formal}
Since the work of Kirkwood \cite{Kirkwood-1935}, it has been appreciated that the free energy difference,
$\df \equiv \df_{0\app 1}$,
 of switching from a Hamiltonian $\ham_0$ to $\ham_1$
is given by a non-linear average,
\begin{equation}
\label{dfcomp}
\df = -k_B T \log{ \left [ \; \langle \, \exp{(-W_{0\app 1} / k_B T) } \, \rangle_0 \; \right ] } \, ,
\end{equation}
where 
$k_B T$ is the thermal unit of energy at temperature $T$ and
$W_{0\app 1}$ is the work required to switch the system from $\ham_0$ to $\ham_1$; see below.
The angled brackets indicate an average over switches starting from configurations drawn from the equilibrium distribution governed by $\ham_0$.

While non-equilibrium approaches to free energy calculations have been a major motivation for this work, we should point out that our analysis applies equally to ``staged'' calculations, in which the free energy is calculated as a sum of increments.
In particular, if one writes the free energy as a sum of incremental parts,
\begin{equation}
\label{dfstage}
\df_{0\app 1} = \df_{0\app\lambda_1} +  \df_{\lambda_1\app\lambda_2}
                + \cdots + \df_{\lambda_k\app1} \, ,
\end{equation}
then each increment $\df_{\lambda_i\app\lambda_j}$ is still defined by a non-linear average analogous to (\ref{dfcomp}) and thus will suffer from finite-sampling error. 

The work $W_{0\app 1}$ required for the average (\ref{dfcomp}) can be defined in a straightforward manner.
In instantaneous switching the work is defined by 
$W_{0\app 1} = \ham_1(\xbf) - \ham_0(\xbf)$ for a start (and end) configuration $\xbf$.
However, gradual switches requiring a ``trajectory''-based work definition may also be used, as was demonstrated by Jarzynski \cite{Jarzynski-1997a,Jarzynski-1997b}.
In this latter case, one requires a Hamiltonian which interpolates between $\ham_0$ and $\ham_1$;
a common choice is
\begin{equation}
\label{hamlam}
\ham(\lambda; \xbf) \equiv \ham_0(\xbf) + \lambda \left[ \ham_1(\xbf) - \ham_0(\xbf) \right] \, ,
\end{equation}
where $\lambda$ is a switching parameter that varies from zero to one.
The work performed in switching gradually from $\ham_0$ to $\ham_1$ along a trajectory $(\lambda(t); \xbf(t))$ is given by
\begin{equation}
\label{work}
W_{0\app 1} = \sum_i \left[ 
  \ham(\lambda_i; \xbf_{i-1}) - \ham(\lambda_{i-1}; \xbf_{i-1})
\right] \, ,
\end{equation} 
where the subscripted configuration $\xbf_{i-1}$ is the (unique) \emph{final} configuration for which $\lambda=\lambda_{i-1}$ --- i.e., the last configuration before $\lambda$ is incremented to $\lambda_i$.
In other words, the work is computed as the sum of those energy increments resulting only from changes in $\lambda$.

Whenever a convex, nonlinear average such as (\ref{dfcomp}) is estimated computationally, that result will \emph{always} be systematically biased \cite{Stone-1982} because one has only a finite amount of data --- say, $N$ work values. 
The bias results from incomplete sampling of the smallest (or most negative) $W_{0\app 1}$ values \cite{Kofke-2001a,Kofke-2001b}:
these values dominate the average (\ref{dfcomp}) and cannot be sampled perfectly for finite $N$, regardless of the $W_{0\app 1}$ distribution.
This is true even for a rectangular distribution;
the sole exception is the trivial $\delta$ function, single-point probability density.
Because of the undersampling of small work values, a running estimate of $\df$ will typically decline as data is gathered, as one sees in the ``staircase'' plots of Fig.\ \ref{fig:examples}.
Such considerations led Wood \emph{et al.} \cite{Wood-1991b} to consider the block-averaged $n$-data-point estimate of the free energy based on $N = m n$ total work values $\{W^{(k)} \}$, namely,
\begin{equation}
\label{dfncomp}
\df_n = \lim_{m \app \infty} \frac{1}{m} \sum_{j=1}^m 
          -k_B T \log{ \left[ \frac{1}{n} \sum_{k=(j-1)n+1}^{jn}
            \exp{ ( -W^{(k)} / k_B T ) } 
                       \right ] } \, .
\end{equation}
It represents the expected value (mean) of a free energy estimate from $n$ data points --- that is, of
\begin{equation}
\label{dfninstance}
\dfnp_n = -k_BT \log{ \left[ \left. 
  \left( e^{-W_1/k_BT} + \cdots + e^{-W_n/k_BT} \right) \right/ n
  \right] } \, , 
\end{equation}
where $m$ estimates have been made.
Wood \emph{et al.} estimated the lowest order correction to 
$\df \equiv \df_{\infty}$ as $\sigma_w^2 / 2 n k_B T$, where $\sigma_w^2$ is the variance in the distribution of work values, $W$ \cite{Wood-1991b}.
Ferrenberg, Landau and Binder discussed analogous issues for the magnetic susceptibility \cite{Ferrenberg-1991,Landau-Binder}.

The derivations below employ continuum expressions simplified by the definitions
\begin{equation}
\label{defs}
w \equiv W/k_BT, \hspace{1cm} 
f \equiv \df/k_BT, \hspace{1cm}
f_n \equiv \df_n/k_BT \, .
\end{equation}
 In terms of the probability density $\rho_w$ of work values, which is normalized by $\int \totd w \rho_w(w) = 1$,
the free energy is given by the continuum analog of (\ref{dfcomp}),
\begin{equation}
\label{dfint}
f = \df/k_BT  = - \log{ \! \left [ \int \!\! \totd w \, \rho_w(w) \, e^{-w}
                   \right ] } \, .
\end{equation}
The form (\ref{dfint}) also occurs in equilibrium calculations \cite{Gubbins-1982}, and forms the basis for the analysis of Lu and Kofke \cite{Kofke-2001a,Kofke-2001b}.
Finally, the finite-data average free energy, following (\ref{dfncomp}) must apply the logarithm ``before'' the average of the $n$ Boltzmann factors, and one has \cite{Zuckerman-2002b}
\begin{eqnarray}
\label{dfnint}
f_n & = & -\int \prod_{i=1}^n \left[ \totd w_i \, \rho_w(w_i) \right] \,
    \log{ \! \left [ \frac{1}{n} \sum_{i=1}^n e^{-w_i}
          \right ] } \, . 
\end{eqnarray}

\section{Generalized Jensen's Inequalities: Block-Averaged Estimates as Rigorous Bounds}
\label{sec:bound}

Jarzynski \cite{Jarzynski-1997b} and subsequently the present authors \cite{Zuckerman-2002a} observed that the free energy is bounded according to
\begin{equation}
\label{bound}
\df \leq \df_n \, , \hspace{0.5cm} \mbox{any } n \, .
\end{equation}
Here we prove this inequality and a generalization originally stated in \cite{Zuckerman-2002a}, namely, 
\begin{equation}
\label{strong-bound}
\df_{n+1} \leq \df_n \, , \hspace{0.5cm} \mbox{any } n \, .
\end{equation}
Indeed the proof given below applies to a broad class of nonlinear averages and functions.
As noted in \cite{Zuckerman-2002b}, the result (\ref{bound}) substantially extends the previous bound $\df \leq \langle W \rangle \equiv \df_1$ \cite{Reinhardt-1992}.
Further, the bounds apply \emph{for an arbitrary distribution of work values} --- that is, whether the probability density of $W$ is multimodal, unimodal, or even rectangular.
We note, finally, that reference \cite{Zuckerman-2002b} by the present authors failed to acknowledge the original statement of the bound (\ref{bound}) by Jarzynski in \cite{Jarzynski-1997b}.

In fact, our proof will show that (\ref{bound}) and (\ref{strong-bound}) are special cases of a more general inequality that depends solely on the convexity and monotonicity of the function used to form a nonlinear average: 
in the case of $\df$ the function is the exponential --- see (\ref{dfcomp});
the root-mean-square is another example, when the function is $g(x) = x^2$.
In the remainder of this section, we use the mathematical convention that upper-case letters (e.g., $X$) indicate random variables whose particular values are specified by lower-case letters (e.g., $x$).

The new bounds are generalizations of Jensen's inequality (see \cite{Hardy-1967}), 
a fundamental property of convex functions with a host of applications including in information theory \cite{Cover-1991}.
Jensen's inequality relates the expected value of a convex function $g$ of a random variable to the same function of the expected value of its argument according to
\begin{equation}
\label{jensen}
\langle g(X) \rangle \geq g(\xavg) \, ,
\end{equation}
where the expectation value is defined in the usual way for an arbitrary function $A$ as
\begin{equation}
\label{expec}
\langle A \rangle = 
\langle A(X) \rangle = \int \!\! {\mathrm{d}} x \, \rho(x) \, A(x) \, ,
\end{equation}
and $\rho$ is the probability density function.
By applying $g^{-1}$ to (\ref{jensen}), the inequality can be re-stated in terms of non-linear and linear averages, respectively,
\begin{equation}
\label{jen-avg}
\xnl \equiv
g^{-1} \left( \langle g(X) \rangle \right) \geq \xavg \, ,
\end{equation} 
with the additional constraint that $g$ be increasing (so that $g^{-1}$ is unique).
Note that the inequality (\ref{jen-avg}) can easily be generalized by applying the inverse of a different increasing function (say, $h^{-1}$) to (\ref{jensen}).


We now state and prove the new ``generalized Jensen's inequalities.''\\
\noindent \emph{Theorem:} \\
Consider estimates for the non-linear average $\xnl$ based on samples of size $n$, 
$\{x_1, x_2, \ldots, x_n\}$,
the expectation of which may be written as
\begin{equation}
\label{xnln}
\xnl_n = 
  \int \!\! {\mathrm{d}}x_1 \, \rho(x_1)  \cdots
  \int \!\! {\mathrm{d}}x_n \, \rho(x_n) \;
  g^{-1} \mbox{\large(} \left. \left[ \,
         g(x_1) + \cdots g(x_n)
         \, \right] \, \right/ n \mbox{\large)} \, .
\end{equation}
Note that 
$\xnl_1 = \xavg$
and
$\xnl_\infty = \xnl$.
Then the new inequalities, generalizing (\ref{jen-avg}), are
\begin{equation}
\label{jen-gen}
\xnl_{n} \geq \xnl_{n-1} \, .
\end{equation}
Strict inequality obtains whenever the random variable $X$ is not limited to a single value (i.e., whenever the probability density $\rho$ is not a single Dirac delta function).
The direction of the inequality is reversed for \emph{decreasing} convex functions, for instance yielding (\ref{bound}) for $g(x) = \exp{(-x/k_BT)}$.

\vspace{0.5cm}
\noindent \emph{Proof:}\\
Note first that $\xnl_n$ is defined in (\ref{xnln}) as the non-linear average based on the ``weighted set'' $S_n$ of all possible $n$-samples $\{x_1, \ldots, x_n\}$.
The weight of each $n$-sample is of course its probability density
$\rho_n(\{x_i\}) = \prod_{i=1}^n \rho(x_i)$.
We will require an explicit construction of the set $S_{n-1}$ from $S_n$, which fortunately is straightforward:
for every $n$-sample with weight $\rho_n$ in $S_n$, if one assigns equal weights $\rho_n/n$ to each of the $n$ available ($n-1$)-samples given by deletion of a single element --- namely,
$\{x_2, x_3, \ldots, x_n\}$,
$\{x_1, x_3, x_4, \ldots, x_n\}$, and so on --- 
one arrives at $S_{n-1}$.
The correctness of this construction follows from the factorizability of the density $\rho_n$, and may be seen by considering the density of a particular ($n-1$)-sample,
$\{\xhat\} = \{\xhat_1, \ldots, \xhat_{n-1} \}$, which can be constructed from $n$ different deletions:
\begin{eqnarray}
\label{setconstruct}
\rho_{n-1}(\{\xhat\}) & = & \frac{1}{n} \left[ 
  \int {\mathrm{d}} x \, \rho_n( x, \xhat_1, \ldots, \xhat_{n-1} )
  + \int {\mathrm{d}} x \, \rho_n( \xhat_1, x, \xhat_2, \ldots, \xhat_{n-1} )
  \right. \nonumber \\
  &&
  \left.
  \hspace*{0.7cm} + \cdots
  + \int {\mathrm{d}} x \, \rho_n( \xhat_1, \ldots, \xhat_{n-1}, x )
  \right]
  \nonumber \\
  & = & \frac{1}{n} \left[
    n \int {\mathrm{d}} x \, \rho(x) \, \prod_{i=1}^{n-1} \rho(\xhat_i)
    \right] \, = \prod_{i=1}^{n-1} \rho(\xhat_i).
\end{eqnarray}

Because of this construction of $S_{n-1}$ from $S_n$, it is sufficient to show that the \emph{single-sample} non-linear average of an arbitrary $n$-sample, namely,
\begin{equation}
\label{udef}
u_n(\{x_i\}) = g^{-1} \left( \frac{1}{n} \sum_{i=1}^n g(x_i) \right) \, 
\end{equation}
exceeds the average $u_{n-1}$ based on the $n$ available ($n-1$)-samples constructed from deletions, as above.
Note that 
\begin{equation}
\label{ytoxnln}
\mbox{\large$\langle$} \, u_n(\{x_i\}) \, \mbox{\large$\rangle$}_n  = \xnl_n \, , 
\end{equation}
which follows immediately from (\ref{xnln}). 

To complete the proof, observe that the single-sample non-linear average (\ref{udef}) can be re-written in terms of smaller samples:
\begin{equation}
\label{ytoless}
u_n(\{x_i\}) = g^{-1} \left( \frac{1}{n} \sum_{j=1}^n
   \frac{1}{n-1}\sum_{i \neq j}^n g(x_i) \right) \, .
\end{equation} 
This identity may be illustrated by considering $g(x_1)$ which occurs $n-1$ times (whenever $j\neq1$), and hence is properly weighted as in (\ref{udef}).
The expression may be further re-written if we denote by
$\{x_i\}_{[j]}$ the original $n$-sample with the $j$th element deleted.
To each of these smaller samples, there corresponds a single-sample, non-linear average $u_{n-1}( \{x_i\}_{[j]} )$.
Applying $g$ to both sides of (\ref{udef}) and substituting the result for $n-1$ into the right-hand-side of (\ref{ytoless}), we then have
\begin{equation}
\label{ytoy}
u_n(\{x_i\}) = g^{-1} \left( \frac{1}{n} \sum_{j=1}^n
  g\!\left[ \, u_{n-1}\mbox{\large(} \{x_i\}_{[j]} \, \mbox{\large)}
    \right]
                      \right) \, .
\end{equation}
If we now consider $U_{n-1} \leftrightarrow u_{n-1}$ to be a random variable with a discrete, $n$-point distribution, we can apply the original non-linear-average inequality (\ref{jen-avg}) to the right-hand side of (\ref{ytoy}), and obtain the desired result
\begin{equation}
\label{yvsy}
u_n(\{x_i\}) \geq \langle u_{n-1} \rangle_{[j]} \, , 
\end{equation} 
where the average $\langle \cdots  \rangle_{[j]}$ is performed over the discrete distribution comprised of all $u_{n-1}$ values obtained from applying (\ref{udef}) to the $n$ sets $\{x_i\}_{[j]}$. 
This completes the proof because when the left-hand-side is averaged as in (\ref{ytoxnln}), the construction of $S_{n-1}$ from $S_n$ guarantees that the average over all $n$-samples on the right-hand side of (\ref{yvsy}) results in $\xnl_{n-1}$ and hence (\ref{jen-gen}). 

The result applies to \emph{any} probability density $\rho$ because no assumptions were made regarding the distribution.
Figure \ref{fig:examples} illustrates the monotonicity of finite-data free energy estimates from two completely unrelated systems.

\section{Asymptotic Behavior: Finite Moments Case}
\label{sec:dfn-finite}

\subsection{Formal Development of the Expansion}

It is possible to generate a formal expansion for the finite-data estimate $f_n$ in terms of $n^{-1}$ for an arbitrary distribution of work values $\rho_w$.
In this section we consider the case where the second and some higher moments of the $z=e^{-w}$ distribution are finite.
Motivated by the central and related limit theorems \cite{Ash-1970,Feller-1971,Uchaikin-Zolotarev} for the sum of the $e^{-w}$ variables, we introduce a change of variables which will permit the development of a $1/n$ expansion for $f_n$.
In particular, we define
\begin{equation}
\label{ydef}
y = ( e^{-w_1} + \cdots + e^{-w_n} - n e^{-f}  ) \, / \, b_1 n^{1/\alpha} \, , 
\end{equation}
where $b_1$ is a constant and $\alpha \leq 2$ is an exponent characterizing the distribution of the variable $e^{-w}$. 
The requirement that $\df$ be finite in (\ref{dfint}) further implies $\alpha > 1$.
The finite-data free energy difference can now be written
\begin{equation}
\label{dfnofy}
f_n 
  = -\int_{-cn^a}^\infty \totd y \, \rho_n(y)  
      \log{ \! \left( e^{-f} + \frac{b_1}{n^a} y \right) } \, , 
\end{equation}
where $c = \exp{(-f)} / b_1$,
$a \equiv (\alpha-1)/\alpha < 1/2$, and $\rho_n$ is the probability density of the variable $y$ normalized appropriately via
\begin{equation}
\label{rhonorm}
\int_{-cn^a}^\infty \totd y \, \rho_n(y) = 1 \, .
\end{equation}
Note that $a$ is always positive because $\alpha>1$.

The expansion of $f_n$ proceeds by first noting that
the sum of \emph{any} set of independent random variables, suitably normalized as in (\ref{ydef}), has a distribution which may be expressed as a stable (\levy) distribution function multiplied by a large-$n$ asymptotic expansion \cite{Feller-1971,Christoph-Wolf}.
In the case of a Gaussian limiting distribution (i.e., $\alpha=2$ or the central limit theorem), assume the variable $z=e^{-w}$ possesses finite ``Boltzmann moments'' --- a mean $\muz=e^{-f}$, 
variance $\sigz^2 = \langle (z-\muz)^2 \rangle$, and 
higher central moments $\muz_p = \langle (z-\muz)^p \rangle$. 
The normalizing constant in (\ref{ydef}) is then $b_1 = \sigz$.
The Boltzmann moments of course differ from the moments of the distribution of $w$.

The so-called Edgeworth corrections to the central limit theorem indicate that the variable $y = (\sum_{i=1}^n e^{-w_i} - n \muz)/\sqrt{n} \sigz$ [cf.\ (\ref{ydef})] is distributed according to \cite{Abramowitz-Stegun,Petrov-1995}
\begin{equation}
\label{cltn}
\rho_n(y) = \rho_G(y;1)
  \left[ 1 + \nu_1(y) / \sqrt{n} + \nu_2(y) / n + \cdots
  \right], \,
\end{equation}
for large $n$, where the remaining terms are higher integer powers of $1/\sqrt{n}$ and the Gaussian density is
\begin{equation}
\label{guass}
\rho_G(y; \sigma) = \exp{(-y^2/2 \sigma^2)} / \sqrt{2 \pi} \sigma \, .
\end{equation}
The functions $\nu_i$, which are defined based upon the Hermite polynomials \cite{Abramowitz-Stegun,Feller-1971}, depend on the original distribution of $e^{-w}$.
In terms of the cumulants $\kap_i$ (see, e.g., \cite{Abramowitz-Stegun}) of the distribution of $z=e^{-w}$ and the Hermite polynomials defined via
\begin{equation}
\label{hermite}
\frac{\totd^k}{\totd x^k} \, \rho_G(x;1) = (-1)^k H_k(x) \, \rho_G(x;1)
\, ,
\end{equation}
the lowest-order Edgeworth functions are \cite{Abramowitz-Stegun,Petrov-1995}
\begin{eqnarray}
\label{nu1}
\nu_1(y) & = & (\kap_3/6 \sigz^3) H_3(y) 
           = (\muz_3/6 \sigz^3) \left(y^3 - 3 y \right)
\\
\nu_2(y) & = & ( \kap_4 / 24 \sigz^4 ) H_4(y)
             + ( \kap_3^2 / 72 \sigz^6 ) H_6(y)
\\
\nu_3(y) & = & ( \kap_5 / 120 \sigz^5 ) H_5(y)
             + ( \kap_3 \kap_4 / 144 \sigz^7 ) H_7(y)
             + ( \kap_3^3 / 1296 \sigz^9 ) H_9(y)
\, .
\end{eqnarray}
The $\nu_i$ functions are odd or even according to whether $i$ is odd or even, in this $\alpha=2$ case.

Before the expansion for $f_n$ can be developed, the integral (\ref{dfnofy}) must be considered carefully by dividing it into three parts:
\begin{eqnarray}
\label{dfn3ints}
- f_n & = &
  \int_{-cn^a}^\infty \totd y \, \rho_n(y)  
      \log{ \! \left( e^{-f} + \frac{b_1}{n^a} y \right) } 
\nonumber
\\
& = & 
  \int_{-cn^a}^\infty \totd y \, \rho_n(y)\log{ \! \left( e^{-f} \right) }
\nonumber 
\\
\label{fnparts}
&&
  + \int_{-cn^a}^{+cn^a} \totd y \, \rho_n(y)\log{ \! \left( 1 + y/cn^a \right) }
  + \int_{+cn^a}^\infty \totd y \, \rho_n(y)\log{ \! \left( 1 + y/cn^a \right) } 
\\
& \equiv & 
  -f + \bigi(-cn^a, cn^a) + \bigi(cn^a,\infty) \, ,
\end{eqnarray}
where the first integral in (\ref{fnparts}) has been evaluated exactly using the normalization of $\rho_n$ (\ref{rhonorm}) and $\bigi$ represents the latter integrals of (\ref{fnparts}).
One can now proceed by using an expansion for the logarithm in 
$\bigi(-cn^a, cn^a)$ and by bounding terms in 
$ \bigi(cn^a,\infty)$.

It is possible to demonstrate rigorously that the second integral in (\ref{fnparts}), $\bigi(cn^a,\infty)$, does not materially contribute to $f_n - f$ for large $n$.
Although, the logarithm cannot be expanded in a power series for
$y > cn^a$,
the integral can be bounded by expressing the log as the integral of its derivative:
\begin{eqnarray}
\label{boundlog}
\bigi(cn^a,\infty)
& = & \int_{+cn^a}^\infty \totd y \, \rho_n(y) 
        \int_1^{1+y/cn^a} \totd x \, x^{-1}
\\
\label{boundviapower}
& \leq & \int_{+cn^a}^\infty \totd y \, \rho_n(y)
        \int_1^{1+y/cn^a} \totd x \, x^{-1+\epsilon}
  = \frac{1}{\epsilon} \int_{+cn^a}^\infty \totd y \, \rho_n(y)
      \left[ \left( 1 + \frac{y}{cn^a} \right)^\epsilon - 1 \right]
\, ,
\end{eqnarray}
with $0 < \epsilon \leq 1$.
To extract the leading behavior of this bound, one can use the expansion of $\rho_n$ (\ref{cltn}) and set $\epsilon = 1$. 
Noting that $a=1/2$, one obtains
\begin{equation}
\label{boundexpand}
\bigi(cn^a,\infty) \leq \frac{1}{c\sqrt{n}}
  \int_{+cn^a}^\infty \totd y \, \rho_G(y;1)
    \left[  
       1 + \nu_1(y) / \sqrt{n} + \nu_2(y) / n + \cdots \,
    \right] \, y \, .
\end{equation}
Using the asymptotic properties of the error function \cite{Abramowitz-Stegun}, one can show that the strongest $n$ dependence of
$\bigi(cn^a,\infty)$
is no stronger than
\begin{equation}
\label{boundlead}
n \exp{( -c^2 n/2 )} \, .
\end{equation}

The leading behavior of $f_n - f$ is thus expected to result from the first non-trivial integral in (\ref{fnparts}), $\bigi(-cn^a, cn^a)$.
Noting again that $a=1/2$ in this case, we may write
\begin{eqnarray}
\label{expandrholog}
\bigi(-c\sqrt{n}, c\sqrt{n})  & = &  
  \int_{-c\sqrt{n}}^{+c\sqrt{n}} \totd y \, 
    \rho_G(y;1)
    \left[ 
       1 + \nu_1(y) / \sqrt{n} + \nu_2(y) / n + \nu_3(y) / n^{3/2} + \cdots \,
    \right] \,
\nonumber \\
&&
    \hspace{3cm} \times
    \left[
       y/c\sqrt{n} - (y/c\sqrt{n})^2/2 + (y/c\sqrt{n})^3/3 - \cdots \,
    \right]
\, .
\end{eqnarray}
What are the leading terms?
There are no terms proportional to $n^{-1/2}$ raised to any odd power because of symmetry considerations:
the $\nu_i$ functions are even for even $i$.
The leading terms are thus \emph{integer powers} of $n^{-1}$, and the expansion of the finite-data free-energy difference is of the form,
\begin{equation}
\label{fnclt}
f_n = f + \varphi_1/n + \varphi_2/n^2 + \cdots \, ,
\end{equation}
where the $\varphi_i$ are constants which depend on the distribution of $z=e^{-w}$.

The explicit correction terms to $f_n - f$ may now be obtained.
First note that asympototic analysis of the integrals appearing in (\ref{expandrholog}) in terms of the error function \cite{Abramowitz-Stegun} indicates that the limits of integration may be extended to $(-\infty,+\infty)$ with errors proportional to 
$\exp{(-c^2n/2)}$.
Straightforward integration then yields the coefficients of the expansion (\ref{fnclt}), namely,
\begin{eqnarray}
\label{phi1}
\varphi_1 & = & \sigz^2 / 2 \muz^2 \, ,
\\
\label{phi2}
\varphi_2 & = & -(4 \muz \muz_3 - 9 \sigz^4) / 12 \muz^4 .
\end{eqnarray}

\subsection{Coefficients for the Gaussian case}
When the distribution of work values is Gaussian, $\rho_w(W) = \rho_G(W,\sigma_w)$, the Boltzmann moments and, hence the $\varphi$ coefficients of (\ref{fnclt}), may be computed analytically.
Note that one \emph{cannot} assume that $z=e^{-w}$ obeys a Gaussian distribution because $z$ is always non-negative. 
The moments follow from straightforward integration, which yields
\begin{equation}
\label{gaussmom}
\left\langle z^p \right\rangle
  = \int \totd W \rho_w(W) e^{-pW/k_BT}
  = \exp{\! \left[ \, p^2 \sigma_w^2 \, / \, 2 (k_BT)^2
         \right]}
\, .
\end{equation}
The $f_n$ expansion coefficients then follow trivially from substitution into (\ref{phi1}) and (\ref{phi2}).
Setting $s=\sigma_w/k_BT$, one finds for the first two coefficents
\begin{eqnarray}
\label{phi1gauss}
\varphi_1 & = &
\left.\left( e^{s^2} - 1 \right) \right/ 2
\, ,
\\
\label{phi2gauss}
\varphi_2 & = &
\left.\left( 
-4 e^{3s^2} + 9 e^{2s^2} - 6 e^{s^2} + 1
\right) \right/ 12
\, .
\end{eqnarray}
To compare this with the finding of Wood \emph{et al.} for $f_n - f$, 
one can expand (\ref{phi1gauss}) for small $\sigma_w$.
One finds
$\varphi_1  \approx \sigma_w^2 / 2 k_BT$, 
which is precisely the first-order prediction of Wood \emph{et al.} \cite{Wood-1991b}.

This analytic calculation explicitly indicates the practical shortcomings of the expansion (\ref{fnclt}).
Although the leading term in $f_n - f$ is linear in $1/n$, the leading coefficients are exponential in the \emph{square} of the distribution's width.
The asymptotic expression (\ref{fnclt}) thus represents a viable approximation only for a very small window about $1/n = 0$ when $s \gg 1$;
see Fig.\ \ref{fig:gauss}.
When asymmetry is added to a Gaussian distribution via the first Edgeworth correction (see (\ref{cltn}) and, e.g., \cite{Feller-1971}), one finds that the exponential dependence of the $\varphi_i$ on $\sigma_w$ is only corrected linearly by the now non-zero third moment of the $W$ distribution.

\subsection{Expansions for the root-mean-square and similar averages} 
The root-mean-square (or standard deviation) is perhaps the best known example of a non-linear average.
The full analysis carried out above carries over quite directly, and indeed applies to any non-linear average.
We will now briefly consider general ``root-mean-powers'' (``power means'').

To be specific, consider the non-linear average resulting from a general power $q=2,4, \ldots$, denoted
\begin{equation}
\label{rmpdef}
\rmpq \equiv \left \langle x^q \, \right \rangle^{1/q} \, ,
\end{equation} 
where $x$ is a variable distributed according to the (arbitrary) probability density $\rho_x$.

In direct analogy with (\ref{dfnint}) one can define the finite-data average for $\rmpq$ as
\begin{equation}
\label{rmpn}
\rmpq_n = \int \prod_{i=1}^n \left[ \totd x_i \, \rho_x(x_i) \right] \,
  \left[ \frac{1}{n} \sum_{i=1}^n x_i^q
  \right]^{1/q} \, .
\end{equation}
The asymptotic expansion follows from the same procedure as above.
One finds that the expansion
\begin{equation}
\label{rmpexpand}
\rmpq_n = \rmpq + \phiq_1/n + \phiq_2/n^2 + \cdots \, , 
\end{equation}
has coefficients
\begin{eqnarray}
\label{rmpcoeff}
\phiq_1 & = &
  (-1/2q)(1-q^{-1}) \rmpq \, \sigq^2 / \muq^2
\\
\phiq_2 & = &
  \left[ q^{-1} (1-q^{-1}) (2-q^{-1}) \rmpq / \muq^4
  \right]
  \left[ \muq \muq_3 / 6 \, - \, (3-q^{-1}) \sigq^4 / 8
  \right] \,
\end{eqnarray}
where $\muq, \, \sigq^2, \mbox{ and } \muq_3$ denote the mean, variance, and third central moment --- respectively --- of the distribution of the variable $x^q$.

\section{Asymptotic Behavior: Divergent Moments Case}
\label{sec:dfn-diverge}
When the distribution $\rho_z$ of the variable $e^{-w} \equiv z$ in (\ref{ydef}) possesses a long-tail,
the limiting distribution is not a Gaussian and the results (\ref{cltn}) and (\ref{fnclt}) no longer hold.
In particular, if one of the tails of $\rho_z(z)$ decays as $z^{-(1+\alpha)}$ with $\alpha < 2$ (implying an infinite Boltzmann variance, $\sigz^2$), then the distribution of the variable $y$ in (\ref{ydef}) approaches a non-Gaussian ``stable'' (\levy) law for large $n$ \cite{Uchaikin-Zolotarev}.   
Note that such power-law behavior in $z$ corresponds to \emph{simple exponential decay in the work distribution.}

A long-tailed $z$ distribution $\rho_z \equiv \rho_1$ also alters the \emph{form} of the asymptotic expansion of the distribution of the sum-variable (\ref{ydef}) and, hence, the expansion of $f_n$ --- which no longer includes solely integer powers of $n^{-1}$, as in (\ref{fnclt}).
Instead of (\ref{cltn}), the $y$ distribution now takes the more complicated form \cite{Christoph-Wolf}
\begin{equation}
\label{stablen}
\rho_n(y) = \rho_\alpha(y) \left[
  1 + {\sum}^* \nu_{uv}(y) / n^{\theta(u,v)} 
  \right] \, , 
\end{equation}
where $\rho_\alpha$ is the appropriate stable probability density with exponent $\alpha$ \cite{Feller-1971,Zolotarev-1986,Uchaikin-Zolotarev}.
The functions $\nu_{uv}$, which are not available analytically, depend on the original distribution of $e^{-w}$ and partial derivatives of the stable distribution.
The exponents are given by
$\theta(u,v) = (u + \alpha v) / \alpha > 0$,
and the summation ${\sum}^*$ includes $u\geq0$ and $v \geq - \lceil u/2 \rceil$, where $\lceil x \rceil$ denotes the integer part of $x$.

To analyze the asymptotic behavior of $f_n$ in this case, the starting point is again equations (\ref{ydef}) - (\ref{rhonorm}), which are fully general.
It is useful to rewrite (\ref{dfnofy}) by scaling the logarithm's argument by the constant $e^{-f}$ and by subtracting zero in the form of the mean of $y$;
one obtains
\begin{equation}
\label{fndiv}
f - f_n =
  \int_{-cn^a}^\infty \totd y \, \rho_n(y)
      \left[ \, \log{ \! \left( 1 + \frac{y}{cn^a} \right) } - \frac{y}{cn^a} \, \right]
\equiv \ihat(-cn^a,\infty)
\, .
\end{equation}
One can now divide up the domain of integration in (\ref{fndiv}) into sub-parts appropriate for expansions of the logarithm of $\rho_n$, in analogy with (\ref{fnparts}).
Because no explicit forms for stable distributions are known in the range $1 < \alpha < 2$ \cite{Uchaikin-Zolotarev}, we will require separate expansions of $\rho_n$ for $|y| \lesssim 1$ and $y \app \pm \infty$ to obtain appropriate convergent behavior.
The required breakdown of the integral is therefore
\begin{equation}
\label{fndivparts}
f - f_n = \ihat(-cn^a,-1) + \ihat(-1,1) + \ihat(1,\infty) \, .
\end{equation}

Each of the integrals in (\ref{fndivparts}) requires a slightly different procedure.
The first, $\ihat(-cn^a,-1)$,
requires an expansion of the logarithm along with the ``short-tail'' $y \app -\infty$ expansion of $\rho_n \approx \rho_\alpha$ (see below).
The second integral, $\ihat(-1,1)$, uses simple convergent series expansions of both the logarithm and $\rho_\alpha$.
Finally, $\ihat(1,\infty)$ requires primarily the ``long-tail'' $y \app \infty$ expansion of $\rho_\alpha$;
the series expansion of the logarithm is also used to show that extending the lower limit of integration to zero accrues a non-leading correction.

Because we will extract only the leading term of $f_n - f$, it is sufficient to use only the leading contribution to $\rho_n$;
that is, considering (\ref{stablen}) we may use the asymptotically valid ($n \app \infty$) approximation
$\rho_n \approx \rho_\alpha$.
(The leading behavior for $f_n$ in the finite-moments case arises, similarly, from $\rho_n \approx \rho_G$.)
The required series expansions for $\rho_\alpha$ in the case of positive summands $z=e^{-w}$ are \cite{Feller-1971,Zolotarev-1986,Uchaikin-Zolotarev}
\begin{eqnarray}
\label{rhoalphaseries}
\rho_\alpha(y;\xi) 
  & = \sum_{k=1}^\infty C_k^0 \; |y|^{k-1}
\, ,
& \hspace{1cm} |y| > 0
\\
  & \approx \sum_{k=1}^\infty C_k^\infty \, y^{-(k\alpha+1)}
\, ,
& \hspace{1cm} y \app \infty
\end{eqnarray}
where the ``$\approx$'' sign denotes an asympotic expansion, and the coefficients --- which depend on the sign of $y$ --- are given by
\begin{eqnarray}
\label{rhoalphacoeff1}
C_k^0(\xi)
  & = & \frac{1}{\pi} 
  (-1)^{k-1} \frac{ \Gamma(1+k/\alpha) }{ k! } 
  \sin{(k \pi \xi/\alpha)} \, , 
\\
\label{rhoalphacoeff2}
C_k^\infty & = &  \frac{1}{\pi} 
  (-1)^{k-1} \frac{ \Gamma( k \alpha + 1 ) }{k!} 
  \sin{(k \pi \xi^+)} \, ,
\end{eqnarray}
with
$\xi^+ \equiv \xi(y>0)=\alpha-1$ and
$\xi^- \equiv \xi(y<0)=1$.
Note that 
$C^0_k(\xi^+) = (-1)^{k-1} C^0_k(\xi^-)$, and in particular,
$C^0_1(\xi^+) = C^0_1(\xi^-) \equiv C^0_1$.
Because the summands considered here are strictly positive, the left tail of the distribution does not exhibit power-law behavior;
rather, it may be termed ``short'' or ``light'' and, asymptotically, is given by \cite{Uchaikin-Zolotarev}
\begin{equation}
\label{light-tail}
\rho_\alpha(y \app -\infty) \approx
  \frac{1}{ \sqrt{ 2 \pi \alpha (\alpha-1) } }
  \left| \frac{y}{\alpha} \right|^\frac{1-\alpha/2}{\alpha-1}
  \exp{ \left\{ - (\alpha-1) 
                  \left| \frac{y}{\alpha} \right|^\frac{\alpha}{\alpha-1}
        \right\} } \, .
\end{equation}

We can now consider the terms in (\ref{fndivparts}) using (\ref{rhoalphaseries}) - (\ref{light-tail}). 
For the sake of brevity we quote only the leading terms, which result from straightforward integrations (after discarding non-leading terms and corrections):
\begin{eqnarray}
\label{intneg}
\ihat(-cn^a,-1) & \approx &
-\frac{  \alpha^2 \, 
        \Gamma \! \left( a+\mbox{$\frac{3}{2}$}, \, \alpha^a (\alpha-1) \right)  }
     {  2 c^2 (\alpha-1)^{a+\frac{1}{2}} \sqrt{ 2 \pi \alpha (\alpha-1) }  }
\;
n^{-2a}
\\
\label{intmed}
\ihat(-1,1) & \approx &
- \left( C_1^0 / 3 c^2 \right) 
\;
n^{-2a}
\\ 
\label{intpos}
\ihat(1,\infty) & \approx &
- \left( C^\infty_1 \idag_\alpha / c^\alpha \right) \, n^{1-\alpha} 
\, ,
\end{eqnarray}
where $2a=2(\alpha-1)/\alpha$, $\Gamma( \cdot, \cdot )$ is the incomplete gamma function \cite{Abramowitz-Stegun}, and 
\begin{equation}
\label{idag}
\idag_\alpha = \int_0^\infty \frac{\totd x}{x^{1+\alpha}} 
                         \left[x - \log{(1+x)} \right] 
         < \infty
\, .
\end{equation}

By comparing powers of $n$ in (\ref{intneg}) - (\ref{intpos}) one sees that the leading behavior of the finite-date free energy estimate, not surprisingly, results from the ``heavy'' power-law tail ($y \app \infty$).
Thus, using (\ref{intpos}), one has
\begin{equation}
\label{dfndiv}
f_n - f \approx \varphi_{\alpha-1} / n^{(\alpha-1)} \, ,
\end{equation}
with 
$\varphi_{\alpha-1} = \left( C^\infty_1 \idag_\alpha / c^\alpha \right) > 0$.
Note that $\varphi_{\alpha-1}$ depends on $\alpha$ and also on the original probability density $\rho_z$ through $c = e^{-f}/b_1$.
Furthermore, one should not expect (\ref{dfndiv}) to be a useful estimate for $f_n -f$:
the next leading exponent, $2(\alpha-1)/\alpha$ is very close to $\alpha-1$ for $\alpha \lesssim 2$. 

\section{Universal Asymptotic Fluctuations}
\label{sec:fluct}
The fluctuations in the finite-data free energy, $f_n = \df_n / k_BT$, as measured by the variance $\sigma_n$ of $\dfnp_n$ of (\ref{dfninstance}), are of considerable interest because of their potential to provide parameter-free extrapolative estimates of $f_\infty = \df/k_BT$ \cite{Zuckerman-2002a};
see also \cite{Meirovitch-1999}.
The variance is given by
\begin{equation}
\label{sigmangen}
\left( \frac{\sigma_n}{k_BT} \right)^2 = 
  \left\langle \left( \frac{ \dfnp_n - \df_n }{ k_BT } \right)^2 \right\rangle
  = \int_{-cn^a}^\infty \totd y \, \rho_n(y) \left[ \log{(1+y/cn^a)} \right]^2
  - ( f_n - f )^2 \, .
\end{equation}
For $n \app \infty$, it was pointed out in \cite{Zuckerman-2002b} that the simple, linear relation between $f_n -f$ and $\sigma_n^2$ was \emph{independent} of the distribution of work values --- that is, \emph{universal}.
Here, we sketch the derivation for the long-tailed case when the second Boltzmann moment diverges.

To calculate the asymptotic behavior of the fluctuations (\ref{sigmangen}) note first that second term $(f_n - f)^2$ is necessarily of higher order than $f_n-f$.
For the crucial integral of (\ref{sigmangen}), one finds 
\begin{eqnarray}
\label{sigmandiv}
\int_{-cn^a}^\infty \totd y \, \rho_n(y) \left[ \log{(1+y/cn^a)} \right]^2
  & \approx & \frac{1}{n^{\alpha-1}} \frac{C^\infty_1}{c^\alpha} \, \iddag \, ,
\\
\label{idbouble}
\iddag_\alpha & = &
          \int_0^\infty \totd u \, \frac{1}{u^{\alpha+1}} \left[ \log{(1+u)} \right]^2 \, .
\end{eqnarray}
Comparing (\ref{sigmangen}) - (\ref{idbouble}) with (\ref{dfndiv}) and (\ref{idag}), we see that as $n \app \infty$
\begin{equation}
\label{sigma-uni}
f_n - f
  \approx
  \frac{\idag_\alpha}{\iddag_\alpha} \left( \frac{\sigma_n}{k_BT} \right)^2 \, .
\end{equation}  
This is a linear relation that depends only on $\alpha$, via the ratio $\idag_\alpha / \iddag_\alpha$, but is otherwise independent of the initital distribution of work values (or Boltzmann factors).
In the limit $\alpha \app 2$, the ratio
$\idag_\alpha / \iddag_\alpha$ approaches $1/2$, which is the finite-Boltzmann-moment result reported in \cite{Zuckerman-2002b}.
Because numerical evaluation of the integral ratio is non-trivial we note that for 
$\alpha=1.25$, 1.5, 1.75, the corresponding values are
$\idag_\alpha / \iddag_\alpha\simeq1.43$, 0.81, 0.61.

Figure \ref{fig:sigma-uni} illustrates the universal behavior for $\alpha=1.5$.
Two integrable distributions were selected to ensure reliable computations.
The ``simple'' or regulated-power-law distribution is defined by
$\rho_{rp}(z) = \alpha' / (1+z)^{\alpha'+1}$, with $\alpha' = \alpha = 1.5$.
The ``power'' distribution is given by
$\rho_p(z) = z_0 / z^\alpha$, with the choice $z_0 = 10^{-4}$.

\section{Summary and Discussion}
\label{sec:conclude}
This report has expanded upon the brief discussion of Ref.\ \cite{Zuckerman-2002b}, giving a general statistical theory describing the systematic error present in free-energy-difference $\df$ estimates based on a finite amount of data ($N$ work values, $W$).
As in \cite{Zuckerman-2002b}, our focus has been on the large-$N$ asymptotic behavior,
motivated by the need to improve extrapolation procedures first explored in \cite{Zuckerman-2002a}.
However, beyond simply giving further details of the derivations of previous results, this report has made transparent the connection to general non-linear averages:
the bounds of Sec.\ \ref{sec:bound}, which generalize Jensen's inequality, explicitly apply to a broad class of nonlinear computations in addition to $\df$ estimates;
and, Sec.\ \ref{sec:dfn-finite} gives asymptotic expansions for geometric averages, such as the root-mean-square.

The universal, asymptotic relation (\ref{sigma-uni}) between the expected value of the biased $\df$ estimate based on $N$ work values ($\df_N$) and the fluctuation in these estimtates ($\sigma_N$) is one of the more striking results.
We have shown here, in Sec.\ \ref{sec:fluct}, that the relation is universal whether or not the second moment of the distribution of Boltzmann factors, $\exp{(-W/k_BT)}$, is finite --- that is, whether or not the central limit theorem applies.
If not, the stable (\levy) distributions come into play, and the relation between $\df_N$ and $\sigma_N$ depends only on the exponent of the limiting stable distribution.

We hope our results will have practical application in the extrapolation process outlined in \cite{Zuckerman-2002a}, which suggested that dramatic increases in computational efficiency may be possible.
In this context, examination of Pad\'e approximants to the asymptotic series, which can be constructed to also exhibit suitable small-$N$ behavior, may prove fruitful.
We believe, finally, that the statistical foundation laid in Ref.\ \cite{Zuckerman-2002b} and here provides a basis for the crucial but non-trivial task of simply \emph{understanding} convergence in estimates of free energy differences and other non-linear averages.

\begin{acknowledgments}
The authors have benefitted greatly from discussions and correspondence with Michael E.\ Fisher, Gerhard Hummer, Chris Jarzynski, David Kofke, Nandou Lu, Hagai Meirovitch, Hirsh Nanda, Lawrence Pratt, Mark Robbins, Attila Szabo, and David Zuckerman.
Marty Ytreberg provided important comments in the final preparation of the manuscript.
We gratefully acknowledge funding provided by the NIH (under grant GM54782 to T.B.W. and NRSA GM20394 to D.M.Z.), the Bard Foundation, and the Department of Physiology.
\end{acknowledgments}


\newpage
\begin{center} FIGURE CAPTIONS \end{center}

\begin{figure}[h]

\begin{center}
\epsfig{file=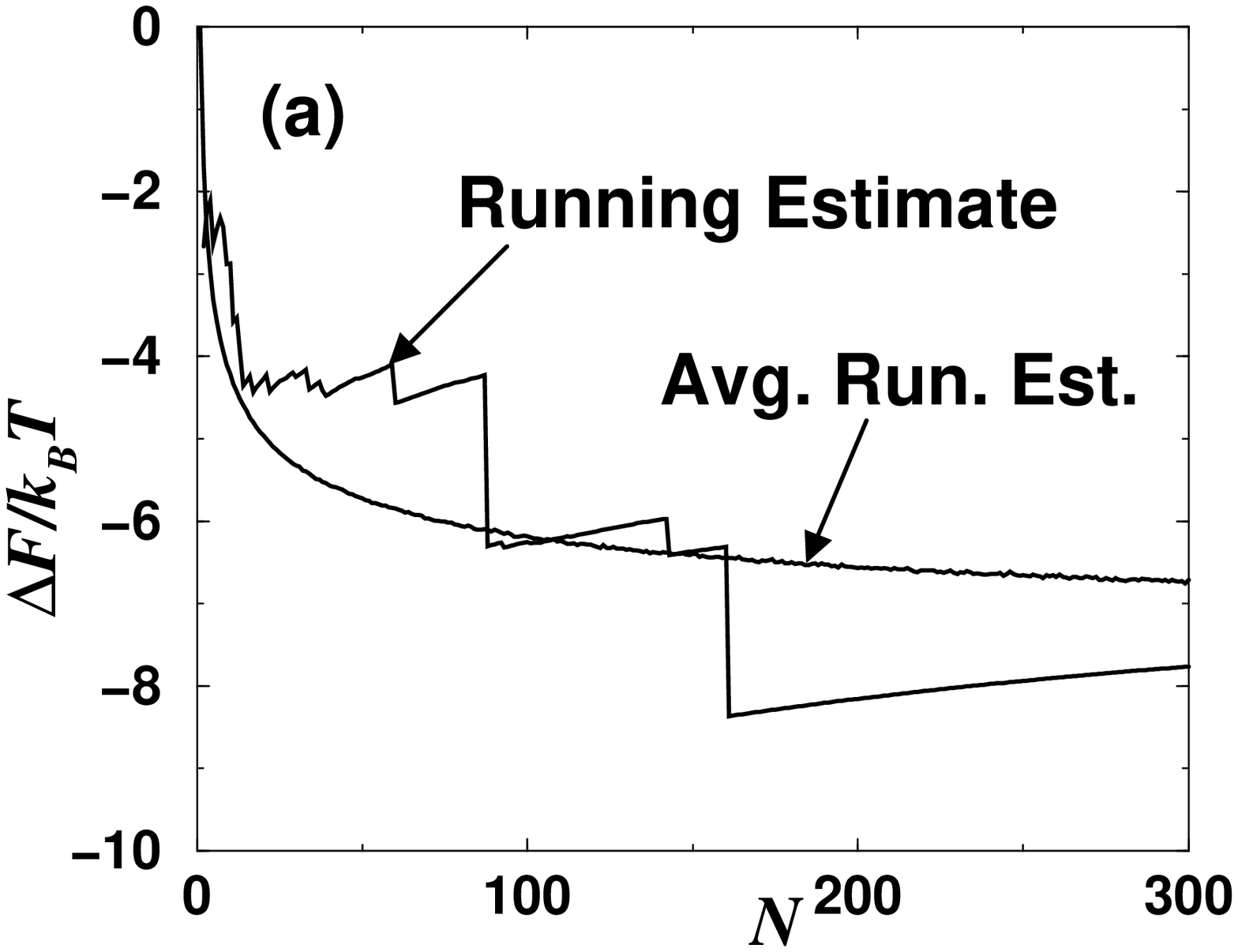,height=2.75in}
\epsfig{file=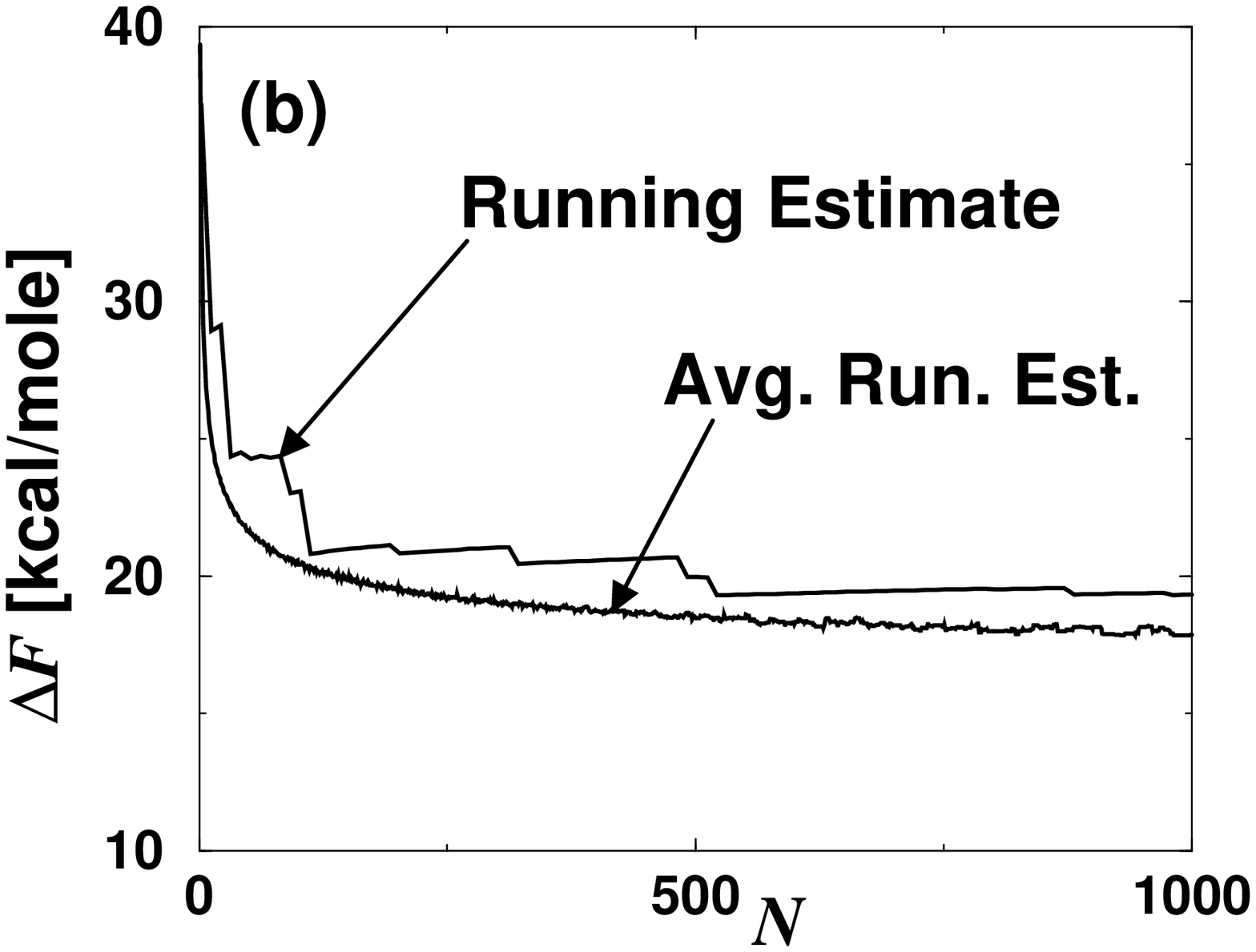,height=2.75in}
\end{center}

\caption{\label{fig:examples}
}
\end{figure}

\vspace{1cm}
\noindent
Figure \ref{fig:examples}.
Finite-sampling errors in $\df$ estimates based on (a) Gaussian-distributed work values and (b) work values generated in a molecular-mechanics solubility comparison between the fatty acids palmitate and stearate.
The irregular, staircase-shaped plots are the running estimates based on $N$ work values, while the smooth curves depict the \emph{average} running estimates $\df_N$ (\ref{dfncomp}) which are independent of the order in which the work values were generated.
The average running estimates are also rigorous upper bounds on $\df$.
The standard deviation of the zero-mean Gaussian distribution in (a) is $4 k_B T$, for which the true free energy difference is $\df = \df_\infty = -8 k_B T$.
For the fatty acid solvation case, $\df = \df_\infty \simeq 13$ kcal/mole;
note that \hbox{1 kcal/mole} = 1.7 $k_BT$.

\begin{figure}[h]

\begin{center}
\epsfig{file=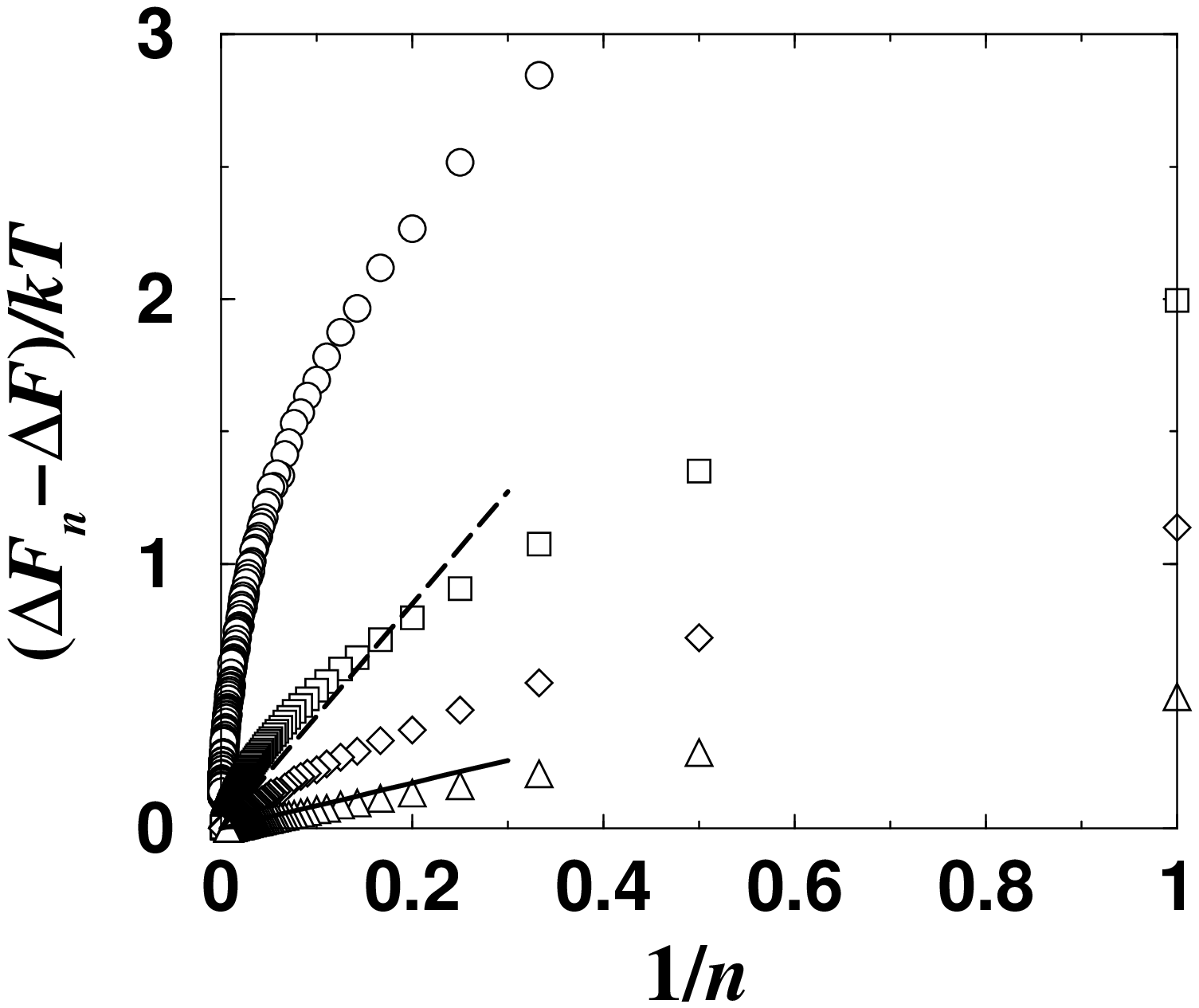,height=2.75in}
\end{center}

\caption{\label{fig:gauss}
}
\end{figure}

\vspace{1cm}
\noindent
Figure \ref{fig:gauss}.
Finite-sampling error for Gaussian-distributed work values.
The expected value of the dimensionless finite-sampling inaccuracy, $(\df_n - \df)/k_BT$ for $n$ data points is plotted as a function of $1/n$.
From top to bottom, the data sets represent numerical values of the error for Gaussian distributions of work values with standard deviations, $\sigma_w/ k_BT$ of 3, 2, 1.5, and 1.
The lines (dashed for $\sigma_w/k_BT = 1.5$, solid for $\sigma_w/k_BT = 1$) depict the asymptotic linear behavior for the two smallest widths.

\begin{figure}[h]

\begin{center}
\epsfig{file=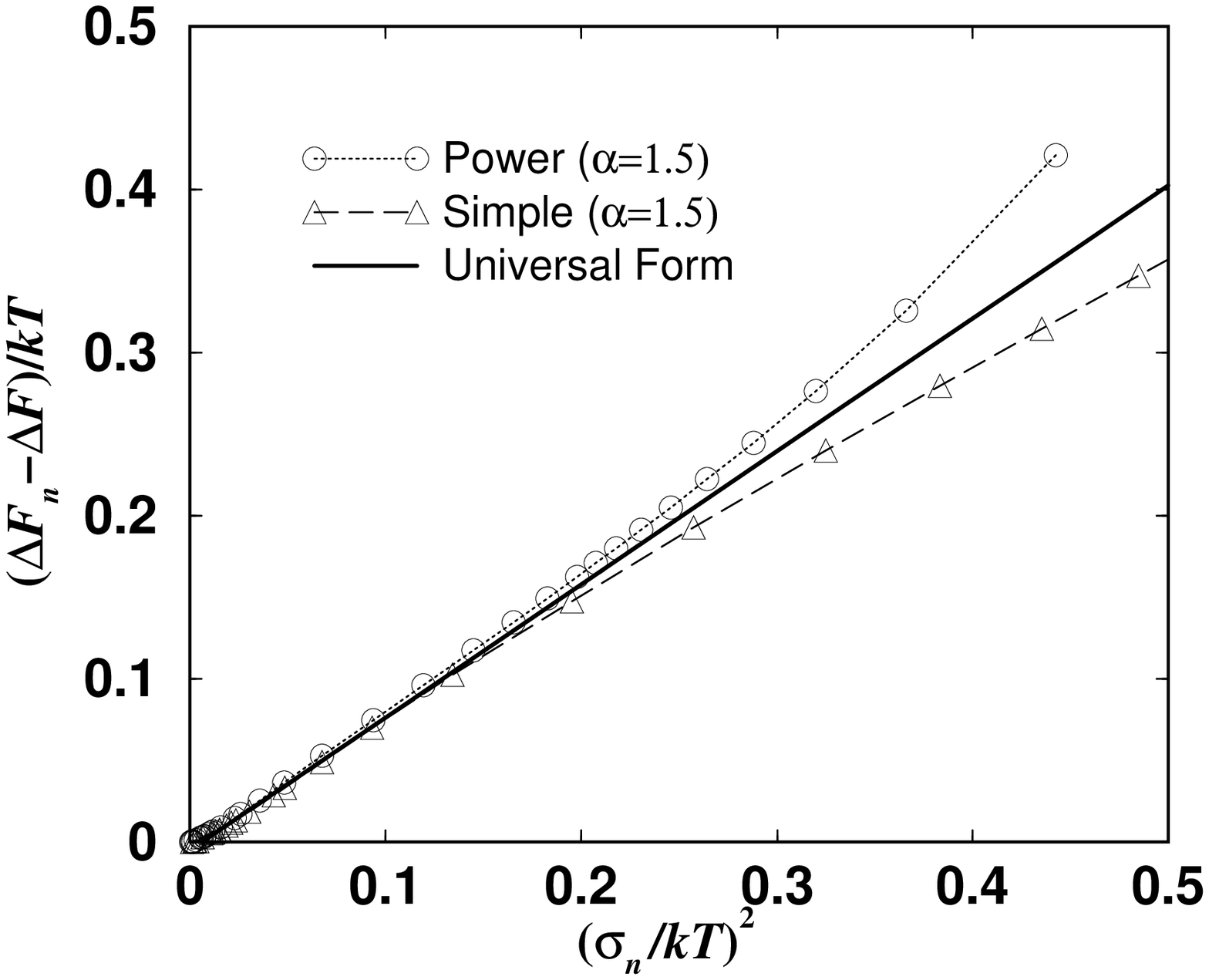,height=2.75in}
\end{center}

\caption{\label{fig:sigma-uni}
}
\end{figure}

\vspace{1cm}
\noindent
Figure \ref{fig:sigma-uni}.
The universal $n \app \infty$ relation between 
$\df_n - \df$ and its fluctuation $\sigma_n$ for the long-tailed case when the \levy\ index is $\alpha=1.5$.
The solid line depicts the universal slope
$\idag_\alpha / \iddag_\alpha \simeq 0.815$ for $\alpha = 1.5$, as given in (\ref{sigma-uni}) and the succeeding text.
The data for the ``power'' and ``simple'' distributions, described in the text, are each shown for $\alpha' = \alpha = 1.5$.

\newpage

\end{document}